\documentclass{article}
\usepackage{spconf,amsmath,graphicx,hyperref}
\usepackage{enumitem}
\usepackage{booktabs}   % 用于支持 \toprule, \midrule, \bottomrule
\usepackage{multirow}   % 用于支持 \multirow
\usepackage{enumitem}     % 用于自定义列表，例如 enumerate 和 description
\usepackage{algorithm}
\usepackage{algorithm}
\usepackage{algpseudocode}
\usepackage{subcaption}    % 提供 subfigure 環境
\usepackage{stfloats}
\usepackage{graphicx}
\usepackage{multirow}
\usepackage{algpseudocode} 
\title{Enhancing Multilingual LLM-based ASR with Mixture of Experts and Dynamic Downsampling}
\name{Guodong Lin$^{1}$, Ziqi Chen$^{1}$, Yuxiang Fu$^{1}$, Ke Li$^{2}$, and Wei-Qiang Zhang$^{1*}$
\thanks{*Corresponding author.}
\thanks{This work was supported by the National Natural Science Foundation of China under Grant No. 62276153.}
\thanks{We gratefully acknowledge the organizers of the MLC-SLM Challenge and
Nexdata for providing the multilingual speech dataset used in this work.}
}
\address{$^1$Department of Electronic Engineering, Tsinghua University, Beijing, China\\
$^2$Beijing Haitian Ruisheng Science Technology Ltd., China\\
{\tt lgd24@mails.tsinghua.edu.cn, wqzhang@tsinghua.edu.cn}}
\begin{document}
%\ninept
%
\maketitle
\begin{abstract}
The rapid progress of large language models (LLMs) has opened up a new frontier for automatic speech recognition (ASR), making their effective integration a critical and challenging research direction. To this end, this work proposes a projector-based LLM-ASR framework targeting the key challenges of multilingual generalization and modality alignment. Our approach incorporates a Mixture of Experts (MoE) architecture to improve cross-lingual adaptability, and a Continuous Integrate-and-Fire (CIF) mechanism for dynamic downsampling and modality alignment. Experimental results show that the combination of these components yields substantial performance improvements, surpassing strong baseline models. The proposed method represents a step toward building more accurate, robust, and generalizable LLM-based ASR systems.
\end{abstract}
\section{Introduction}
\label{sec:intro}

Recent advances in large language models (LLMs) has opened up a new frontier for Automatic Speech Recognition (ASR). Unlike traditional language models that primarily provide statistical sequence constraints, LLMs offer strong semantic reasoning and contextual understanding, making them promising candidates for enhancing ASR systems. 
However, integrating LLMs into ASR systems to fully leverage these capabilities remains a significant challenge.

Early studies have investigated using LLMs as auxiliary modules for ASR system. For example, Hrinchuk et al. \cite{Hrinchuk-postprocess} and Ma et al. \cite{Ma-LLM-postprocess} leveraged LLMs as post-processors to enhance the fluency and correctness of ASR outputs. Other studies \cite{velikovich18_interspeech-semantic-LM-fusion, Li-LLM-fusion, Ma-LM-fusion, hono-etal-2024-integrating-LLM-fusion} treated LLMs as external language models, interpolating LLM decoder scores with ASR outputs, partially enhancing linguistic coherence. However, these approaches still treat the ASR model as the central component and make limited use of the rich contextual knowledge embedded in LLMs.

Recent work has moved toward deeper integration, restructuring ASR under the Encoder–Projector–LLM paradigm (hereafter LLM-ASR). In this framework, a speech encoder produces acoustic representations that are mapped into the LLM embedding space by a projector, enabling the LLM to generate transcriptions. Ma et al.~\cite{ma2024embarrassinglysimpleapproachllm} showed that even with frozen encoder and LLM, a simple trainable linear projector can achieve state-of-the-art performance on LibriSpeech~\cite{librispeech} at low training cost. However, later studies~\cite{eval-of-slamasr, zhang25t_interspeech} reported major limitations of LLM-ASR, including weak cross-domain robustness and hallucination issues, indicating that naive LLM integration is insufficient for reliable ASR.

Building on these insights, two challenges are critical for advancing LLM-ASR. The first is multilingual generalization. Although LLMs naturally support multiple languages, aligning speech encoders with multilingual LLMs requires a flexible mapping capable of adapting to diverse phonetic and linguistic patterns. Mixture-of-Experts (MoE) architectures~\cite{lepikhin2021gshard_moe} offer a promising solution by dynamically routing inputs to specialized experts, and have demonstrated strong scalability and adaptability across languages, accents, and domains~\cite{Streaming_asr_moe, ye24_interspeech, mosa, boosting}. Motivated by these advantages, we introduce an MoE-based projector into the LLM-ASR framework to better model cross-lingual acoustic-to-text mappings.

Another challenge lies in modality alignment, i.e., mapping variable-length acoustic features to text tokens. Existing projector designs~\cite{ma2024embarrassinglysimpleapproachllm, LLM-ASR_yu} typically adopt fixed downsampling for efficiency, but this rigidity makes performance highly sensitive to the downsampling rate~\cite{zhang25t_interspeech}. Moreover, unlike traditional ASR models using CTC~\cite{graves2006connectionist} or attention-based approaches~\cite{attention}, fixed downsampling lacks explicit alignment modeling, leading to degraded robustness under speech rate variations. To address this issue, we adopt Continuous Integrate-and-Fire (CIF)~\cite{cif} for dynamic downsampling. CIF provides lightweight audio–text alignment, enabling the projector output length to adapt flexibly to token sequences and thereby improving modality alignment in LLM-ASR.

In this work, we propose a novel projector-based LLM-ASR integration framework, focusing on two key aspects: improving multilingual speech recognition and enhancing modality alignment for LLM-ASR architecture. Our main contributions are as follows:

\begin{enumerate}
\item \textbf{MoE-Enhanced Projector}: We introduce an MoE-based projector into the LLM-ASR architecture. Through gated routing, the model dynamically selects expert networks, achieving superior multilingual ASR performance compared to Whisper \cite{whisper} and improving cross-lingual generalization and recognition accuracy.

\item \textbf{CIF-Based Modality Alignment}: We replace the projector’s fixed downsampling with a CIF-based dynamic mechanism, which predicts the number of target text tokens and accordingly aggregates audio features. This enables more precise modality alignment and yields substantial gains in both accuracy and robustness.
\end{enumerate}

\section{Methods}
\label{sec:methods}
In this section, we will first introduce the baseline framework that forms the foundation of our work, and then present two key components introduced in this work: the MoE-Enhanced Projector and the CIF-Based Modality Alignment. These two components jointly improve the performance and generalizability of the LLM-ASR architecture, with notable benefits for multilingual ASR.

\subsection{Baseline Framework}
\begin{figure}[htb]
  \centering
  \includegraphics[width=\linewidth]{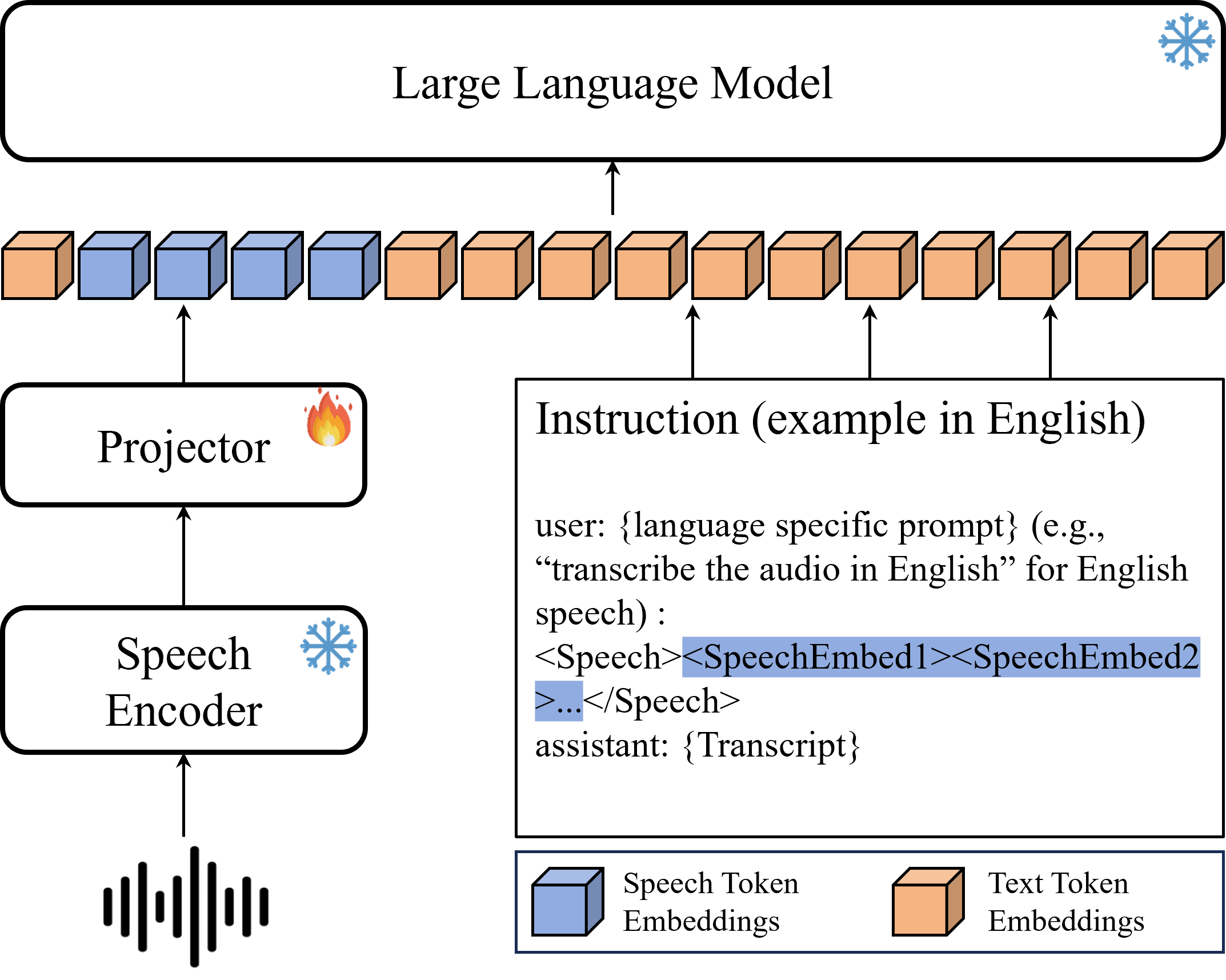}
  \caption{Baseline Framework.}
  \label{fig:baseline}
\end{figure}

Fig. \ref{fig:baseline} illustrates the baseline architecture used in this work, based on the widely used LLM-ASR framework. For implementation, we adopt the open-sourced MLC-SLM-Baseline codebase available on GitHub \footnote{https://github.com/mubingshen/MLC-SLM-Baseline}.

\vspace{-0.5em}
\subsubsection{Speech Encoder}
The speech encoder extracts acoustic features from audio input. Typically, a large pre-trained speech encoder is used for this task. 

\vspace{-0.5em}
\subsubsection{Projector}
The projector module bridges the speech encoder and the large language model, mapping the acoustic features into a token sequence that the LLM can process. This module typically uses a lightweight architecture, such as linear layers.

\vspace{-0.5em}
\subsubsection{Large Language Model}
In this LLM-ASR architecture, the LLM serves as the core decoding module, leveraging its strong capabilities in syntax, semantics, and contextual understanding to perform transcription. To reduce the overall training cost, the LLM is kept frozen during system training.
In our experiments, we further observed that designing language-specific prompts for each language leads to noticeably faster convergence than using trainable prompts (as shown in Fig. \ref{fig:baseline}) or applying a single universal prompt in one language (e.g., English) to all input utterances regardless of their original language. Due to space constraints, detailed results are omitted.

\subsection{MoE-Enhanced Projector}
Although the baseline design, a lightweight stack of two convolutional layers followed by two linear layers, offers efficiency, it is insufficient to handle the complex audio-to-text mapping required in multilingual ASR. To address this limitation, we enhance the projector with an MoE architecture, enabling more effective modeling of diverse linguistic patterns across languages.

Specifically, the original convolutional layers are retained as a shared backbone, and a set of expert sub-networks is added, with the number of experts matching the number of training languages. A gating mechanism then computes an activation weight for each expert sub-network based on the acoustic features. The output of the MoE Projector, which combines the outputs of the selected experts, is given by:
\begin{equation}
\mathbf{y} = \sum_{k=1}^Kg_k(\mathbf{x})E_k(\mathbf{x})\text{,}
\end{equation}
where $g_k(\mathbf{x})$ represents the activation weight assigned by the gating netword for the $k$-th expert sub-network for input $\mathbf{x}$, and $E_k(\mathbf{x})$ denotes the corresponding expert output.

% To address the issue of parameter explosion when the number of languages increases, we also experimented with a lightweight Adapter-based MoE architecture. This variant uses adapters for each expert sub-network to reduce the computational cost while maintaining the flexibility of the MoE model.

\subsection{CIF-Based Modality Alignment}
The encoder typically outputs acoustic feature frames at a frame rate of 50–100 frames per second, while the discrete token sequence required by the LLM has a much lower temporal resolution. To bridge this mismatch, the projector usually applies a fixed downsampling strategy that compresses multiple frames into a single information-dense token. However, such fixed-rate downsampling often fails to align the lengths of audio and text sequences, particularly when speech rates vary substantially across datasets.

To solve this issue, we employ a CIF-Based Downsampling method. CIF is a lightweight module that assigns a weight (typically between 0 and 1) to each acoustic feature frame. These weights are accumulated sequentially across frames, and whenever the cumulative sum exceeds a threshold (e.g., 1), the predictor fires, indicating that a new token has been detected. A well-trained CIF predictor can estimate the number of text tokens corresponding to an utterance and compress the acoustic feature sequence to a length that closely matches the target token sequence. This dynamic downsampling enables more accurate alignment between the projector output and the target token sequence.

In our experiments, CIF-based downsampling often yields a substantially higher compression rate than fixed-rate approaches, where the baseline framework typically applies a downsampling factor of 4. Such over-compression may lead to information loss and degraded recognition performance. To mitigate this issue, we adopt a relaxed CIF variant by modifying the training objective to encourage the predictor to generate an output sequence that is $n$ times longer than the target token sequence. With an appropriate choice of $n$, this approach achieves an effective downsampling rate comparable to  the baseline, while retaining CIF’s adaptability to variations in speech rate.

\begin{figure}[htb]
  \centering
  \includegraphics[width=\linewidth]{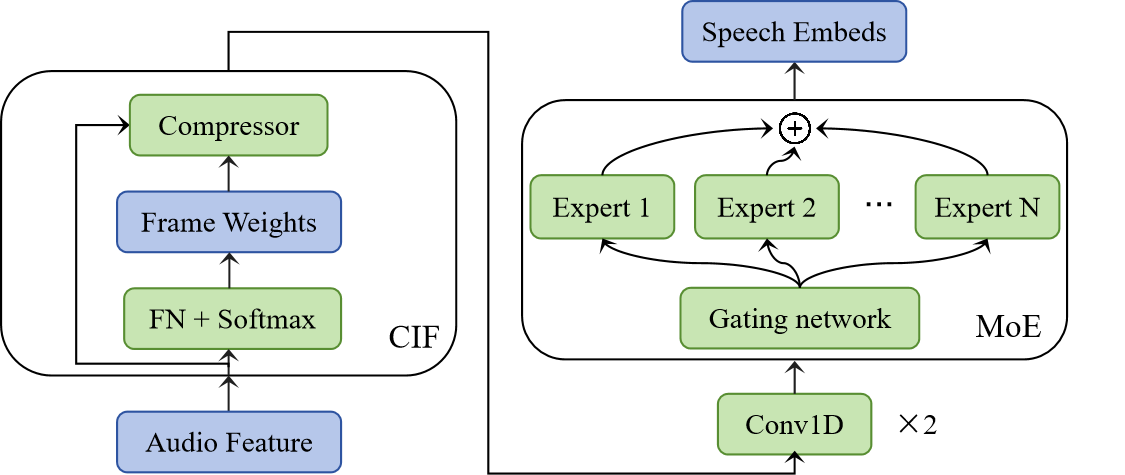}
  \caption{Structure of the projector used in our approach.}
  \label{fig:projector}
\end{figure}

As shown in Fig. \ref{fig:projector}, the integration of the MoE-enhanced projector with CIF-based modality alignment enables the system to more effectively map speech to text, improving both accuracy and robustness.

\section{Experiment Setup}
\label{sec:experiment}
\subsection{Models and Modules}
\subsubsection{LLM-ASR Backbone Architecture}
In all experiments, we use the Whisper-large-v3 encoder (hereafter referred to as Whisper-encoder) as the speech encoder and Qwen-2.5 7B as the backbone LLM.
For the projector module, we select a baseline version consisting of two convolutional layers and two linear layers, with a fixed downsampling factor of 4. The MoE-Enhanced Projector is then derived by replacing the linear layers with MoE layers.

\subsubsection{CIF Predictor}

The CIF predictor consists of a single convolutional layer followed by a linear layer. It is first trained jointly with a frozen Whisper-encoder, where the reference transcripts are tokenized using the Qwen-2.5 7B tokenizer. The training objective is the mean squared error (MSE) between the predicted output length of the CIF predictor and the target length. For the original CIF, the target length corresponds to the reference token sequence length, while for the modified CIF, it is set to $n$ times this length. In our experiments, we set $n=4$, which yields an effective downsampling rate comparable to that of the baseline system with a fixed downsampling factor of 4.

After pretraining, the CIF predictor is inserted into the predictor, replacing the original fixed-ratio downsampling component. During subsequent training, the CIF predictor remains frozen.

\subsection{Datasets}
We use the 1,500-hour multilingual ASR dataset provided by Nexdata for the MLC-SLM Challenge\footnote{https://www.nexdata.ai/competition/mlc-slm} to train both the projector and CIF modules. The dataset covers 11 languages: English, French, German, Italian, Japanese, Korean, Portuguese, Russian, Spanish, Thai, and Vietnamese. English contributes 500 hours across five accents, while each remaining language provides 100 hours.

As the ground-truth transcriptions of the MLCSLM test set are not publicly available, the development set is used as the in-domain (ID) evaluation benchmark. For models trained on 1,500 hours, out-of-domain (OOD) evaluation is conducted on the FLEURS \cite{fleurs} and CommonVoice \cite{commonvoice} test sets in the same 11 languages.

To study the impact of data scaling on performance and generalization, we further conduct data expansion experiments for the proposed method by augmenting the original MLC-SLM training set with an additional 6,500 hours of speech randomly sampled from CommonVoice, GigaSpeech2 \cite{gigaspeech2}, LibriSpeech, MLS \cite{MLS}, and VoxPopuli \cite{voxpopuli}, resulting in a total of 8,000 training hours. Under this setting, MLCSLM-dev and CommonVoice-test are treated as ID benchmarks, while FLEURS-test serves as the sole OOD evaluation.

% To further investigate the scaling behavior of our architecture with respect to training data volume and to enhance generalization ability, we additionally incorporate several open-source datasets, including Common Voice, GigaSpeech2, KsponSpeech, LibriSpeech, ReazonSpeech, MLS, TED-LIUM3, VoxPopuli, and Yodas. To maintain a balanced distribution across languages, we apply language-dependent random sampling ratios. The final training set amounts to a total of 8000 hours, with per-language distributions shown in Table X.

% After incorporating additional open-source training data, we further introduced the Common Voice and FLEURS test sets to evaluate the model’s performance on in-domain (ID) and out-of-domain (OOD) conditions, respectively.

% \begin{table*}[t!]
% \centering
% \caption{Main performance comparison between our proposed method and baselines. }
% \label{tab:main_comparison}
% \resizebox{\textwidth}{!}{%
% \begin{tabular}{p{0.3\textwidth}|c|c c}
% \toprule
% Method & Trainable Params & WER (\%) on MLCSLM-dev  & WER (\%) on FLEURS-test  \\
% \midrule
% LLM-ASR Baseline & 37.73M & 23.26 & 13.05 \\
% Whisper-large-v3 & / & 21.48 & 9.59 \\
% \midrule
% LLM-ASR + MoE projector & 133.14M & 16.10 & 11.06 \\
% \textbf{Proposed}  & 75.09M & \textbf{15.27} & 10.31 \\
% \bottomrule
% \end{tabular}%
% }
% \end{table*}
\begin{table*}[t!]
\centering
\caption{Average WER (\%) on three evaluation datasets.
For 1{,}500 h training, MLCSLM-dev is in-domain, with FLEURS-test and CommonVoice-test as out-of-domain; for 8{,}000 h training, MLCSLM-dev and CommonVoice-test are in-domain, and FLEURS-test is out-of-domain.
Results include Whisper-large-v3, the baseline LLM-ASR, and its MoE- and CIF-based variants.
}
\label{tab:main_comparison}
\resizebox{\textwidth}{!}{%
\begin{tabular}{l|c|c c c}
\toprule
Method & Training data duration (h) & MLCSLM-dev  & CommonVoice-test & FLEURS-test \\
\midrule
Whisper-large-v3 & - & 21.48   &  12.53 & 9.59\\
\midrule
LLM-ASR Baseline & 1500 & 23.26  & 19.57 & 13.05\\
\quad + MoE Projector  & 1500 &16.10  & 14.48 &  11.06\\
\quad \quad + CIF Downsampler & 1500 & 18.95  & 18.45 & 12.89\\
\multirow{2}{*}{\textbf{Proposed (MoE + modified CIF)}}
& 1500 & \textbf{15.27}  & \textbf{13.87} & \textbf{10.46}\\
& 8000 & \textbf{15.45} & \textbf{9.86} & \textbf{8.65} \\
\bottomrule
\end{tabular}%
}
\end{table*}

\subsection{Training Details}
All experiments are conducted on six NVIDIA A40 GPUs, each with 48 GB of memory. For optimization, we employ the Adam optimizer with a maximum learning rate of 1e-3. The learning rate is linearly warmed up over the first 2,500 steps and subsequently decayed according to an inverse square root schedule.
Training is performed for up to 20 epochs, with early stopping applied when the validation loss no longer decreases. To mitigate fluctuations in GPU memory usage due to variable input lengths, we employ a dynamic batching strategy, constraining each batch to a maximum of 5,000 frames.

\subsection{Experiments}
We use Whisper-large-v3 and the original LLM-ASR architecture as performance baselines.
On top of the LLM-ASR framework, we progressively incorporate the MoE-Enhanced Projector and the CIF-Based Modality Alignment, evaluating each component individually to verify its effectiveness.

\subsection{Results}
\label{sec:results}

As shown in Table \ref{tab:main_comparison}, the baseline LLM-ASR model performs worse than Whisper-large-v3 on all datasets, highlighting the difficulty of extending this architecture to multilingual ASR.

\begin{figure}[htb]
  \centering
  \includegraphics[width=\linewidth]{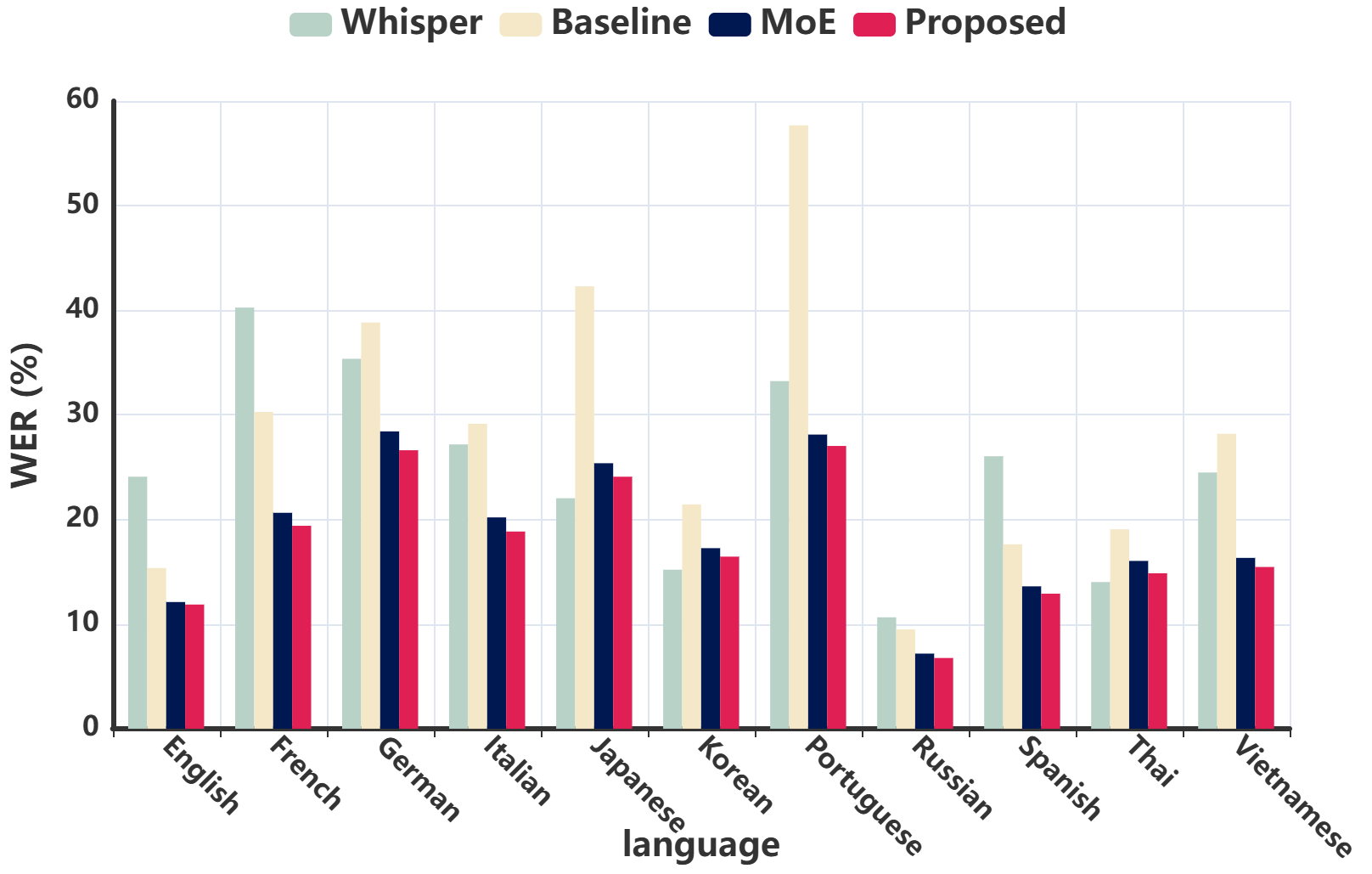}
  \caption{WER by language on MLCSLM-dev dataset.}
  \label{fig:mlcslm_lang}
  \vspace{-1em}
\end{figure}

The MoE-Enhanced Projector significantly improves performance, lowering the WER on MLCSLM-dev to 16.10\%. The per-language breakdown in Fig.~\ref{fig:mlcslm_lang} shows that MoE consistently enhances recognition across most languages. This confirms the effectiveness of MoE in handling diverse cross-lingual mappings.

However, replacing the fixed downsampler with the standard CIF leads to degraded performance (18.95\% on MLCSLM-dev), consistent with previous reports that CIF’s aggregation often induces overly aggressive compression and information loss. By adopting the modified CIF, our proposed method achieves 15.27\% on MLCSLM-dev, while also attaining the best results on both FLEURS (10.46\%) and CommonVoice (13.87\%).  After scaling to 8{,}000 hours, the WER on MLCSLM-dev shows a slight increase, while substantial gains are observed on CommonVoice-test and FLEURS-test, which were originally or remain out-of-domain. This suggests that large-scale data expansion primarily enhances cross-domain generalization, with a mild trade-off on the tightly matched in-domain benchmark. These results demonstrate that the modified CIF retains the adaptability of CIF while maintaining an effective downsampling rate, yielding the strongest overall performance across all datasets.
\section{Conclusion}
This paper introduces two key enhancements to the LLM-ASR framework: MoE-based projector and CIF-based dynamic downsampling mechanism. These improvements significantly boost system performance, highlighting the importance of well-designed projectors and alignment strategies for robust, generalizable LLM-ASR systems. Future work will explore advanced modality alignment strategies.
% In this paper, we present a novel LLM-ASR framework that tackles the critical challenges of multilingual generalization and modality alignment. We introduced two key architectural enhancements: an MoE-based projector and a CIF-based dynamic downsampling mechanism. These improvements significantly enhance the performance of the LLM-ASR system, demonstrating that carefully designed projectors and alignment strategies are essential for building robust and generalizable LLM-based ASR systems. Future work will explore advanced modality alignment strategies.

{\small
\bibliographystyle{IEEEtran}
\bibliography{refs}
}

\end{document}